\documentclass[fleqn,twoside]{article}
\usepackage{espcrc2}

\usepackage{graphicx}

\newcommand{\AmS}{{\protect\the\textfont2
  A\kern-.1667em\lower.5ex\hbox{M}\kern-.125emS}}

\title{
{
\vspace{-3.0cm} \normalsize \hfill
\parbox{30mm}{HU-EP-04/48\\SFB/CPP-04-36\\[3mm] September 2004}
}\\[15mm]
Cutoff--effects in the spectrum of dynamical Wilson fermions}

\author{M. Della Morte\address[HU]{Institut f\"ur Physik, Humboldt Universit\"at, 
        Newtonstr. 15, 12489 Berlin, Germany},
        R. Hoffmann\addressmark\thanks{Talk presented by R. Hoffmann at Lattice 2004},
        F. Knechtli\addressmark,
        and        
        U. Wolff\addressmark[HU]}
       
\begin{document}
\thispagestyle{empty}
\begin{abstract}
\vspace{-0.1pc}
We investigate the low--lying eigenvalues of the
improved Wilson--Dirac operator in the 
Schr\"odinger functional with two dynamical quark
flavors. At a lattice spacing of approximately 0.1 fm
we find more very small eigenvalues than in the
quenched case. These cause problems with HMC--type
algorithms and in the evaluation of fermionic correlation
functions. Through a simulation at a finer lattice spacing
we are able to establish their nature as
cutoff--effects.
\vspace{-0.4pc}
\end{abstract}

\maketitle

\section{Motivation}

Recently more and more evidence has been accumulated that for
dynamical improved Wilson fermions  at a lattice
spacing of $a\!\simeq\!0.1\rm fm$ the cutoff--effects are much larger
than expected from quenched experience (for a summary of large
scaling violations in the two--flavor--theory see
ref.~\cite{Sommer:2003ne}). As an extreme
example, for three flavors the existence of a phase transition in the
$\beta$--$\kappa$--plane has been numerically conjectured and is
interpreted as a lattice artifact \cite{Aoki:2001xq}.

In addition several collaborations have reported algorithmic difficulties with
(improved) Wilson fermions on relatively coarse lattices (see e.g.
\cite{Joo:2000dh}), which seem to be related to small eigenvalues
of the Dirac operator. Through the inversion of the Dirac operator
these small eigenvalues can result in large
driving force during the molecular dynamics evolution
of Hybrid Monte Carlo (HMC) algorithms. In turn, the large forces
are likely to trigger instabilities of the numeric integrator employed, which
produces large Hamiltonian violations
and can result in long periods of rejection \cite{DellaMorte:2004hs}.
In this way small
eigenvalues affect algorithmic performance.

The occurrence of very small eigenvalues generates not only algorithmic
problems. On the configurations in question the quark propagator becomes
large and one observes ''spikes'' in fermionic correlation functions.
This affects not only their mean value but also their autocorrelation
function, making the statistical analysis difficult. This is discussed
in more detail in ref.~\cite{DellaMorte:2004hs}.

For dynamical simulations this is an unexpected problem since
na\"{\i}vely it is assumed that the determinant should suppress small
eigenvalues compared to the quenched situation.
In our setup we have two infrared cutoffs (finite quark mass and
Schr\"odinger functional boundary conditions) that should prevent
the Dirac operator from developing very small eigenvalues.
Nevertheless we observe them in simulations with a lattice spacing
of about $0.1\,\rm fm$.
Here we show that these small eigenvalues disappear if one goes
to smaller lattice spacings and can thus be interpreted as a lattice
artifact.

\section{Setup and error analysis}

We simulate the Schr\"odinger functional (SF) with
two dynamical flavors of non--perturbatively improved Wilson
fermions. The algorithms used are HMC with two pseudo--fermion
fields \cite{Hasenbusch:2001ne} and PHMC \cite{Frezzotti:1997ym}.
In the following the term 'eigenvalue' always refers to the
eigenvalue (in lattice units)
of the square of the Hermitian even--odd--preconditioned Wilson--Dirac
operator in the normalization of \cite{DellaMorte:2003jj}.
For PHMC the parameters of the polynomial are chosen
such that more configurations with small eigenvalues are produced
compared to the QCD Boltzmann weight.
After reweighting this gives a very good estimate of the path--integral
weight of such configurations. Another benefit of using PHMC is that
the polynomial provides a regularized inversion, thus also
addressing the algorithmic problems due to large forces mentioned
above. With this algorithm the correct ensemble average of an estimator
$O$ is given by\\[-6mm]
\begin{equation}
\langle O\rangle=\frac{\langle OW\rangle_P}{\langle W\rangle_P},
\end{equation}
\vspace*{-2mm}
where $W$ is the reweighting factor \cite{Frezzotti:1997ym}
and the subscript $P$ indicates
an average over the PHMC--generated ensemble.
HMC--type algorithms are expensive and generate strongly autocorrelated
data, which makes a careful data analysis indispensible.
We use an explicit integration of the
autocorrelation function as described in ref. \cite{Wolff:2003sm}.

To obtain a histogram for a quantity $f\!=\!\langle\phi\rangle$ we
analyze $P_n\!=\!\langle\chi_n(\phi)\rangle$, where $\chi_n$ is the
characteristic function of the $n$-th bin. This also provides us
with an error on the population of the bin, thus giving us the ability to
assess differences between histograms. If the PHMC algorithm is used,
the bin population is given by
\begin{equation}
P_n=\langle\chi_n(\phi)\rangle=\frac{\langle\chi_n(\phi)W\rangle_P}
{\langle W\rangle_P}.\label{phmc}
\end{equation}
\section{Comparison to the quenched case}

In order to test the na\"{\i}ve expectation that the fermionic
determinant suppresses small eigenvalues we compare quenched
and dynamical ensembles of $8^3\!\times\!18$ lattices at roughly
matched physical parameters. Using the quenched data from
ref.~\cite{Garden:1999fg} and the dynamical data
from refs.~\cite{Aoki:2002uc} and \cite{Allton:2001sk}
we cohose the parameters such that the lattice spacing and the
(large volume) pseudo--scalar mass are matched.

\begin{figure}[!t]
\begin{center}
\hspace*{-1mm}\includegraphics[width=77mm]{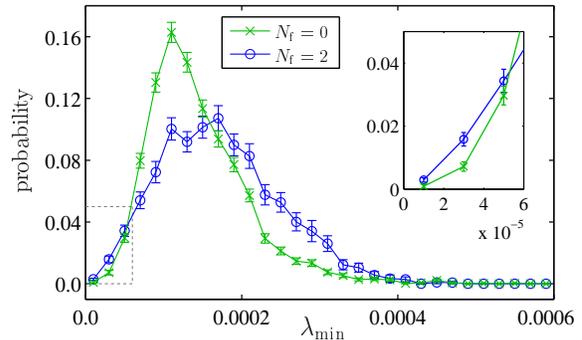}
\vspace*{-14mm}
\caption{Distributions of $\lambda_{\rm min}$ from matched quenched
($\beta\!=\!6.0$) and two--flavor dynamical simulations ($\beta\!=\!5.2$).
.\label{cutoff}}
\vspace*{-12mm}
\end{center}
\end{figure}

The distribution of the smallest eigenvalue $\lambda_{\rm min}$ is
shown in Figure \ref{cutoff}. The quite small error bars we obtain
from eq.~\ref{phmc} at the lower end of the spectrum are due to the
enhanced occurrence of small eigenvalues by PHMC.
While $\langle\lambda_{\rm min}\rangle$
is increased from $1.44(1)\!\cdot\!10^{-4}$ to $1.72(5)\!\cdot\!10^{-4}$
with two dynamical flavors we see that in the infrared tail the dynamical
data show more events. More precisely, the probability of finding
a smallest eigenvalue below $4\!\cdot\!10^{-5}$ increases from
$0.81(16)\%$ to $1.88(26)\%$. To show that this long tail towards
zero is a cutoff--effect we now compare dynamical data from different
lattice spacings at matched parameters.

\section{Finer lattices}

\begin{figure}[!t]
\begin{center}
\vspace*{-0mm}
\hspace*{-1.5mm}\includegraphics[width=78mm]{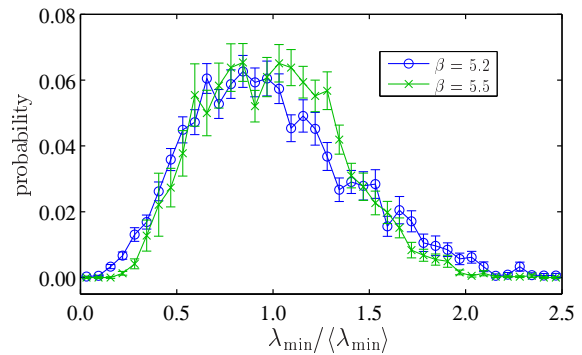}
\vspace*{-12mm}
\caption{Distributions of the smallest eigenvalue from two--flavor
simulations at two different lattice spacings.
\label{roughmatch}}
\vspace*{-11mm}
\end{center}
\end{figure}

Through measurements of the SF coupling $\bar g^2$
\cite{Luscher:1992an,Bode:2001jv} we found that increasing
$\beta$ from 5.2 to 5.5 changes the lattice spacing by
roughly a factor of $2/3$. Anticipating that the algorithmic
problems due to very small eigenvalues would no longer
be present at this finer lattice spacings we used HMC
with two pseudo--fermions to generate an ensemble of
$12^3\times27$ lattices. Ignoring small changes in
the renormalization factor we compare this to an $8^3\times18$
PHMC ensemble at $\beta\!=\!5.2$, which is matched using
$Lm_{\rm PCAC}$, the box size times
the bare PCAC mass. To plot both lattice spacings simultaneously we
divide by $\langle\lambda_{\rm min}\rangle$ in Figure \ref{roughmatch}.

Going from $\beta=5.2$ to $5.5$ reduces the variance (normalized
by the mean value)
of
the smallest eigenvalue from $0.178(10)$ to $0.127(19)$. 
The long tail at the infrared has disappeared at the finer lattice spacing
and we thus interpret it as a cutoff--effect.

In a final step we compare the dynamical data at $\beta\!=\!5.5$ to another
quenched run at approximately the same lattice spacing ($\beta\!=\!6.26$),
volume and bare quark mass.

\begin{figure}[!t]
\begin{center}
\vspace*{0mm}
\hspace*{-1mm}\includegraphics[width=78mm]{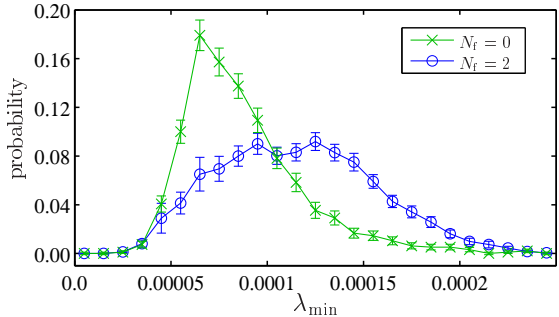}
\vspace*{-15mm}
\caption{Comparison of the smallest eigenvalue distribution at a lattice
spacing of approximately $0.07\,\rm fm$.
\label{matchfine}}
\vspace*{-11mm}
\end{center}
\end{figure}

In Figure \ref{matchfine} both the quenched and the dynamical data show
very similar behavior at the infrared end of the spectrum.
A comparison to Figure \ref{cutoff} shows
that the excess of very small eigenvalues we found at the coarser
lattice spacing has disappeared entirely. As at $a\simeq0.1\,\rm fm$ the
average smallest eigenvalue is still shifted upwards by the fermionic
determinant.

\vspace*{-2.2mm}

\section{Conclusions}

\vspace*{-1.2mm}

The problems in simulating dynamical Wilson fermions at a
lattice spacing of approximately $0.1\,\rm fm$ are due to
the occurrence of very small eigenvalues in the spectrum
of the Wilson--Dirac operator. We use PHMC to better sample
this part of the spectrum in a comparison of two--flavor
and quenched simulations at matched physical
parameters.

As expected we find
the \emph{average} smallest eigenvalue
to be larger in the dynamical case. However, due to its increased
variance the dynamical data shows more
\emph{very} small eigenvalues despite the finite quark
mass and the cutoff provided by the Dirichlet boundary conditions.

At a lattice spacing of approximately $0.07\,\rm fm$
both the dynamical and a matched quenched run show a
spectrum that is well separated from zero. As before
$\langle\lambda_{\rm min}\rangle$ is larger for two flavors.

We conclude that at $\beta\!=\!5.2$, corresponding to a
lattice spacing of $0.1\,\rm fm$, the spectrum of the
Wilson--Dirac operator is strongly distorted. Our simulations show
that this is a cutoff--effect that disappears rapidly
with increasing $\beta$.

We do not expect that these findings
are specific to the Schr\"odinger functional. Without the
additional infrared cutoff due to the boundary conditions one
should see these problems already at larger quark masses.

\vspace*{2mm}

\textbf{Acknowledgements:} We thank R. Sommer for valuable discussions. This work
was supported by the Deutsche Forschungsgemeinschaft (SFB/TR 09) and the
Graduiertenkolleg GK271.

\end{document}